# Deep Learning Detection of Inaccurate Smart Electricity Meters: A Case Study

Ming Liu[§,a,b,c], Dongpeng Liu[§,c], Guangyu Sun[c], Yi Zhao[c], Duolin Wang[c], Fangxing Liu[e], Xiang Fang[d], Qing He[e], Dong Xu[c*]

a. Key Laboratory for Applied Statistics of MOE and School of Mathematics and Statistics, Northeast Normal University, Changchun, 130024, China.

b. Department of Applied Mathematics, School of Mathematics and Statistics Sciences, Changchun University of Technology, Changchun, 130012, China.

c. Department of Electrical Engineering and Computer Science, Christopher S. Bond Life Sciences Center, University of Missouri, Columbia, Missouri 65211, USA.

d. Qinghu Rising Sunshine Data Technology (Beijing) Co., Ltd, Beijing 100084, China;

e. Division of Electromagnetic Metrology, National Institute Ltd, Beijing 100084, China;

§ *Equal contribution. Author ordering determined by coin flip.*

*Abstract*— **Detecting inaccurate smart meters and targeting them for replacement can save significant resources. For this purpose, a novel deep-learning method was developed based on long short-term memory (LSTM) and a modified convolutional neural network (CNN) to predict electricity usage trajectories based on historical data. From the significant difference between the predicted trajectory and the observed one, the meters that cannot measure electricity accurately are located. In a case study, a proof of principle was demonstrated in detecting inaccurate meters with high accuracy for practical usage to prevent unnecessary replacement and increase the service life span of smart meters.**

## I. INTRODUCTION

### A. Rationale

SMART meters are important and popular electrical metering equipment used in the construction of a smart grid [1] [2]. Almost two billion smart meters will be installed worldwide by 2020, and the smart grid will serve nearly 80% of the world's population according to the Navigant Research Report [3]. The report projected that by 2020 China alone is expected to have more than 435 million smart meters installed compared to 132 million for the United States [3]. Smart meters can have various malfunctions; in particular, their measurements can be inaccurate and these meters need to be replaced to ensure the measurement accuracy. The smart meters in China are mandatorily replaced after eight years of service, but most of them are still in good conditions at that time. If inaccurate smart meters can be replaced as needed in a targeted fashion, a huge amount of money and human resources will be saved. For example, if the smart meters' average service life span increases from 8 to 12 years using the replace-as-needed approach, it could save about 12 million dollars per year in China alone (assuming each smart meter is 60 dollars). However, manually checking all smart meters on a regular basis is labor-intensive. Therefore, it is more feasible to apply an intelligent system using electricity measurement data to identify inaccurate smart meters.

*B. Previous Studies on Smart Meter Analyses*

With the ever-increasing broad applications of smart meters, massive data of smart meters are collected every moment. The meters' big data provide a good source for data analyses of smart meters' performance. To detect inaccurate smart meters, one approach is to collect historical data before the malfunction and then model it for predicting the malfunction. A number of related studies have been conducted along this line. Based on the Smart Metering Electricity Customer Behavior Trials (CBTs) data conducted by the Irish Commission for Energy Regulation (CER), Silipo et al. [4] used a K-means algorithm and an auto-regressive model to predict the future usage of energy. However, their approach was not applied to malfunction detection. Similarly, Cosmo et al. [5] used the same data set to explore the relationship between electricity consumption and time-of-use tariffs.

A number of studies targeted malfunction detection of smart meters. Wu et al. [6] implemented a

Shewhart-CUSUM method to identify smart meter malfunction using data in the summer season only. Zufferey et al. [7] introduced artificial neural networks to predict short-term load and photovoltaic of smart meter profiles. By analyzing the malfunction of electric power systems' big data, Sheng et al. [8] developed a big data modeling method for identifying abnormal data based on matrix theory. He [9] analyzed smart meters' historical abnormal data to study the malfunction occurrence probability and malfunction type of smart meters, and he also used neural networks to predict a malfunction. Yang et al. [10] proposed a smart meter malfunction identification model based on abnormality analysis. Yao et al. [11] proposed three reliability prediction methods to estimate smart meter lifetimes based on big data and fault classification. Gajowniczek et al. [12] proposed an approach to predict the electricity load for the next 24 hours.

*C. Previous Studies on Malfunction Detection*

Although these methods are valuable, they have room to improve. None of them reached a good accuracy for large-scale usage. No study has applied any cutting-edge deep learning method for smart meter malfunction detection, even though deep learning methods have been successfully used for several other malfunction detection problems in recent years. For example, Nikovski et al. [13] proposed a predictive model for losses in electrical distribution networks to detect power theft. Zhang et al. [14] employed a deep learning method to detect traffic accidents from social media data. Kim and Cho [15] proposed a method combined with long short-term memory (LSTM) network, a convolutional neural network (CNN), and a deep neural network (DNN) to extract complex features. These studies motivated us to apply deep learning for smart meter malfunction detection, as presented in this paper.

*D. Paper Outline*

This paper focuses on detecting inaccurate smart meters, instead of other types of malfunctions or electricity theft detection. It is arranged as follows: Section II introduces methods for data collection and data cleaning, as well as detection and diagnosis strategies, whereas LSTM and CNN methods are used to build inaccurate smart meter detection models. Section III describes and analyzes the performance, as well as shows experiments and results. Finally, Section IV provides conclusions.

## II. METHOD

The formulation and workflow of the method used in this paper are shown in Fig. 1. In this system, the current, voltage and power usage of every single user are recorded by a submeter, while the total current, voltage and power usage of a residential area (e.g. an apartment) are recorded by the master meter. A few submeters may be out of order, where the meter reading has a measurement error rate above a given threshold (e.g. 10%). These submeters are referred to as inaccurate submeters. The master meter, however, is under routine maintenance and can be assumed accurate within a negligible error rate. The system collects data from all submeters and the master meter. Then, the task can be formulated as a malfunction detection problem to identify inaccurate submeters if any by using electricity usage data. For the malfunction detection, the method used in this paper adopts LSTM to predict the reading of the master meter based on the data collected from submeters and historical data. Then the predicted value is compared with the actual master meter reading. If the predicted value is significantly different from the actual master meter reading for a period of time (determined by a devised metric, named lag to be introduced in Section II D 2), the detection task will predict the presence of inaccurate meters in the residential area, which causes the diagnostic work to be activated. At this time, a TS-RP (Times Series-Recurrent Plot) CNN model is used to identify and classify every submeter to predict whether it is accurate.

### A. Data Collection and Description

First, the data used in this study was collected from two residential areas. The smart meter being studied has five technical specifications: 1) a rated power of 1100 W, 2) a rated voltage of 220 V, 3) a rated current of 5 A, 4) a rated frequency of 50 Hz, and 5) an error rate of 2%. The data is unique compared to previous studies. For example, Silipo et al. [4] predicted electricity usage for different groups of meter IDs, using a data set based on questionnaires completed by users, but they did not record instantaneous current and voltage values, which are included in this paper's analysis. In addition, unlike their data set, the data set used in this study not only records submeter data but also master meter data for a specific residential area. Therefore, their data cannot be used in the problem formulated here.

After concatenating all the data and renaming the columns, the data sets of current and usage (including master meters and submeters) were collected. The master meter has three phases and submeters have only one phase. For a single submeter, three values are collected: voltage, current and electricity usage. The voltage and current tables have 102 columns and 3038 rows each. The voltage and current were collected every 15 minutes in the master meter and every hour in the submeters, which are real-time measurements. The 102 columns of voltage and current tables include region ID, user ID, system ID, phase, multiplier, date, and 4×24 real-time data. Meanwhile, the everyday usage table has six columns and 65,534 rows. The 6 columns of everyday usage include number (from 1 to 65,534), region ID, user ID, system ID, date, and usage. Each submeter has a unique user ID and system ID, and the user ID is chosen as the primary key of this data set.

### B. Data Cleaning and Preprocessing

There are redundant data and invalid data in the data sets. If a single ID's recording (a row in dataset), at the same date and time, appears more than once, then only the first recording is used, and others are regarded as redundant. The invalid data were removed as well, whose Sum of Submeters at each day, named SSub, is larger than the master meter's measurement, which is unrealistic since the submeters measure the electricity consumption of the subbranches from the master meter and other power losses between the master meter and the submeters. Upon removing these data, the time feature was formatted, for example, the weekday, month and year, with respect to electricity usage. A date is converted into a one-hot encoding in 22 (7 weekdays+12 months+3 years) dimensions including the weekday, month and year.

*C. Data Analysis*

The data source will be introduced, the metric that is used in the system, and the data distribution. The desensitized data were collected from two residential areas called Hua Yuan (residential area A) and Dong Hui (residential area B). Residential areas A and B collected the voltage readings and the electric current readings of 104 submeters every hour from August 2014 to August 2016. In addition, the master meters of residential area A and B recorded real-time voltage and current every 15 minutes. All the data in this paper were synchronized in corresponding records after data cleaning and preprocessing.

After the data cleaning, the formula to calculate the measurement residual error (including the transmission power loss) for one day is shown below:

$$E = W_{master} - \sum_{i=1}^{n} W_{sub_i} \qquad (1)$$

where $E$ represents the daily "residual error" between the master meter and SSub, $W_{master}$ represents the daily reading of the master meter, and $W_{sub_i}$ represents the reading of i-th submeter over n submeters at that day. In the raw data, the errors between the master meter and submeters are small, most of the relative errors do not exceed 2%, which indicates high accuracy of collected data. In addition, these meters are new at the time of data collection; so, it is assumed that there is no inaccurate meter.

With the data distribution, one possible approach to detect malfunction is to check and compare the residual error. The residual error distribution of different times roughly follows a normal distribution; however, the mean and variance of distributions vary in different seasons, as shown in Fig. 2. It means the usage changing factors (such as weather changes, acquisition of new appliances, change in consumption patterns, etc.) are mixed up with potential measurement errors from the submeters. Therefore, one cannot simply use the residual error distribution to detect malfunction. Inspired by [16], which provided a successful example for neural network methods to detect an abnormality, in this study various deep neural network models were explored to identify significant measurement errors from smart meters in a robust way. The method actually does not intend to predict electricity usage, but rather to detect abnormalities in the residual error independent from various usage changing factors.

*D. Detection and Diagnosis of Malfunction*

The overall workflow includes **prediction task for residual error, detection task for malfunction injection in a residential area,** and **classification task for inaccurate submeters**, as shown in Fig. 1. Two efficient deep neural networks, LSTM and CNN, are respectively applied to predict the presence (injection) of malfunction in the smart meters inside a residential area, then classify which smart meter is inaccurate. A sliding window detection procedure is proposed to detect the malfunction injection in a residential area. LSTM [17] is a popular deep learning architecture for prediction in time-series data [18][19]. Benefiting from the gated recurrent architecture, LSTM is able to discover both short- and long-range relationships inside given time-series data, which is an essential capability for the residential area's error prediction task. CNNs are widely used in image recognition and sequence tasks like natural language processing [20]. CNN produces a strong response to a spatially local pattern when the convolution kernels capture specific 1D or 2D information of feature maps. In the sequence classification task, the abnormal patterns are distinguished from normal patterns by the classification of CNN. Also, because of shared weights in the architecture, CNN can save significant memory and time in training and prediction, fulfilling the efficiency needs raised by processing a large number of submeters.

*1) Residual Error Prediction Based on LSTM*

In order to predict the residual error vs. time, a two-layer LSTM is used. The input is the features of the residential area including the historical data of residual "error" (float, the error between SSub and the master meter), "master" (float, the measurement of the master meter), "com_date"(int, the time interval between the current date and the base date), "weekday" (list, 7-dimension one-hot code), "month" (list, 12-dimension one-hot code), "year" (list, 3-dimension one-hot code, years 1, 2, 3 in this case), and "number" (int, the number of submeters in the current residential area). The output is the predicted residual error in the future. The architecture of LSTM is described as Fig. 1. Both LSTM-1 and LSTM-2 have 30 dimensions, followed by a dense layer. LSTM calculates a hidden state $h_t$ as follows:

$$f_t = \sigma(x_t U^f + h_{t-1} W^f) \tag{2}$$

$$o_t = \sigma(x_t U^o + h_{t-1} W^o) \tag{3}$$

$$\tilde{C}_t = \tanh(x_t U^g + h_{t-1} W^g) \tag{4}$$

$$C_t = \sigma(f_t * C_{t-1} + i_t * \tilde{C}_t) \tag{5}$$

$$h_t = \tanh(C_t) * o_t \tag{6}$$

where: $i, f,$ and $o$ are the input, forget and output gates, respectively. $W$ is the recurrent connection between the previous hidden layer and the current hidden layer. $U$ is the weight matrix that connects the input to the current hidden layer. $\tilde{C}$ is a "candidate" hidden state that is computed based on the current input and the previous hidden state. $C$ is the internal memory of the unit.

*2) Sliding Window Detection Procedure*

The prediction process aims to tell if there are inaccurate meters in the residential area for the future days by using a sliding-window detection procedure described in Fig. 1 with two predefined parameters, threshold *t* and window size *L*. The residual error between the predicted value and the observed value at a certain day is referred to as the daily prediction error (DPE). In each step, if not all the DPEs in the window exceed the threshold, the window moves forward with a one-day stride. Keep sliding until all DPEs inside the window exceeds the threshold t, possible inaccurate meters are considered in this residential area. If no such a window is found, it is assumed that there is no inaccurate meter in this residential area.

To evaluate the prediction performance, commonly used metrics for a regression task such as RMSE does not reflect the prediction purpose of this study. The devised evaluation metric in the experiment is called "lag," which is defined as the length between the predicted starting date of the malfunction and the actual starting date of the malfunction. When all DPEs inside the window exceed the threshold, the left edge of the window is considered as the predicted starting date of malfunction.

*3) Classification of Submeters Based on TS-RP CNN*

The detection of inaccurate meters is formulated as an inaccurate submeter classification task for each submeter (accurate vs. inaccurate). In this task, 1D-CNN with sequential input and 2D-CNN (VGG16) with matrix input are implemented and compared. Then, the novel TS-RP CNN is built by merging the two types of input.

Much like [21], a 1D-CNN architecture is applied to classify the sequence data, as shown in the bottom half of Fig. 3. To fully utilize the information in the dataset, such as the frequency features, recurrent plots (RP) is applied in the CNN architecture, similar to [22]. RP is a tool for nonlinear data analysis. It visualizes a square matrix, whose elements correspond to those times when a state of a dynamical system recurs [23]. The main advantage of RP is that it can show the phase-space transformation information of the sequential data, even when the data is short-term [24], which is believed to be complementary to the time-series feature. The RP conversion is implemented by using the class library SciPy [25]. Transfer learning [26] is used from VGG16 pre-trained by the ImageNet dataset [27]. A dense layer is added at the end to obtain the 1D output. As a combination of these two models, the TS-RP CNN has two types of input, where the matrix input (RP) is sent to the VGG model and the sequential input is fed into the 1D-CNN model. Then, the two outputs are merged after a dense layer to get the final result. The combined signal is activated by a sigmoid function after the element-wise addition. At first, the sequential input is replaced with the matrix input. The accuracy of the result improves but still cannot have a good area under the curve (AUC) enclosed by the receiver operating characteristic (ROC) curve. Because both sequential input and matrix input enclose various features, these two inputs are combined with two paths as shown in Fig. 3. By comparison with a single-input model, the accuracy of the combined inputs greatly improved. This shows that the sequential input and the PR phase-space input assist each other in the model (that is also why the adding operation outperformed concatenation). Phase-space features have been examined by several researchers. For example, a study [28] recently used fast Fourier transform as a layer in a deep neural network, where a 1D sequence input was converted to 2D phase features using multi-frequency decomposition (MFD). However, they neglected the information inside the original sequential data, while TS-RP CNN utilizes the sequential data as a parallel input. So far, the model is the first to combine time series and corresponding RP as the dual input branched for CNN. Thus, the method is called a time-series (TS) and RP (recurrence plot) CNN, or TS-RP CNN method. The networks in this study are trained with Keras [29] with TensorFlow [30] backend on a Nvidia Geforce 1080ti GPU. The code and example datasets are available online at https://github.com/minoriwww/MeterDetection.

III. EXPERIMENTS AND RESULT

*A. Data Generation for Inaccurate Submeter*

The data in residential area A was thoroughly inspected by humans and no submeter has shown an abnormal reading as all meters were new. Thus, it is assumed that both the submeters and the master meter were accurate. To simulate a real-world residential area with inaccurate meters, the malfunction was embedded on purpose, i.e. a certain offset into a random number of submeter readings. The offset for an inaccurate submeter is a shift in the reading value (in comparison with the actual value) with respect to time, according to the technical specifications of smart meters. The relationship of shift and time can be constant, linear or even exponential. In this work, it is considered linear with random noises.

To generate inaccurate submeter data (negative samples), 30% of submeters were randomly selected with random starting times for malfunction and injected the malfunction factor into every selected submeter's daily usage, the start date. It is worthwhile mentioning that 30% of submeters were chosen for the purpose of method validation. The method can be used for detecting a single malfunctional submeter. The injection process is expressed as:

$$Usage_{new(i)} = (1 + \alpha \cdot (i - s)) \cdot Usage(i) + N, i \geq s \tag{7}$$

where $i$ represents the date sequence, $\alpha$ represents a restriction of error percentage per day, which is set to 0.01 in this paper, and N represents the random noise. For proof of principle, this formula only represents a likely scenario of the error trend based on our experience. The 1% drift per day is not based on real-world data and may be too large, but our model should be able to detect smaller drift in a longer time frame when the total drift reaches a certain threshold in a similar way. To reflect realistic but challenging user scenarios, it is assumed that the injected error to be restricted to a much lower amount than the typical error between SSub and the reading of the master meter; otherwise, the error would be too obvious to ignore by any method. By injecting the malfunction factor into a submeter, recurrence plots for the corresponding submeter and the residual error (from Equation 1) are changed accordingly, while the reading of the master meter remains unmodified.

The final generated data contains a certain proportion of positive and negative samples by sampling. Those data will be used in the **detection task for malfunction injection** and the **classification task for inaccurate submeters**. The former requires modeling based on whether the residential area contains inaccurate submeters by detecting the abnormality in the residual error trajectory between SSub values of the residential area and the master meter values vs. time. The latter requires the model to identify which submeters have malfunction injections. When applying to real-world applications, this system is expected to work seamlessly.

## B. Result of Residual Errors Prediction Based on LSTM

As described in the LSTM model in Section II D (1), a sequence of known master meter residual errors with a certain window (number of days) is used for training to predicting the master meter residual errors of the following days. By shifting the windows, the residual errors can be predicted over time. An example of predicting master meter residual errors is shown in Fig. 4(a). Most of the predicted values closely trace the observed values. Meanwhile, to select the most effective window size for training the LSTM model, different window sizes were tested and the result shows that time window sizes from 40 (days) to 80 (days) had lower mean square errors (MSEs) between the predicted values and observed values, which is less than 0.005, and a comparatively lower standard deviation, as shown in Fig. 4(b). Herein, the window size for the following experiments is set to 40 days. Then, the time series data was divided into sample 1 (day 1 to day 40), sample 2 (day 2 to day 41)…, etc. and 730 such samples were generated. The 730 samples were split into a training set of 703 samples and a testing set of the remaining 27 samples. It is worth mentioning that although there are some overlapping days between training and testing samples, such a setting reflects real-world applications and can be considered independent between training and testing samples from the machine-learning perspective.

For the **malfunction injection detection task**, predicted residual errors by LSTM is compared with the observed residual error $E$ in Equation 1 in the test set. If $E$ overflows the range between the Upper Bound (UB) and Lower Bound (LB) consistently according to Equations 8 and 9, the residential area is regarded as including one or more inaccurate submeters:

$$UB = p + t \tag{8}$$

$$LB = p - t \tag{9}$$

where p represents the predicted result of LSTM and t represents the threshold of the procedure.

Figure 5(a) shows a detection example when the residential area has a malfunction generated as described in Section III A. The parameters of the procedure are set to t = 0.5 and L = 4. It detects malfunction with a lag of 65 days. Figure 5(b) shows a test case of a residential area where no inaccurate submeter was detected under the same parameter settings. If the residential area had malfunctioning meter(s), the predicted and measured values were sent to the **inaccurate submeter classification task**.

*C. Result of Malfunction Submeter Classification Based on TS-RP CNN*

The receiver operating characteristic (ROC) curve and precision-recall curve (PRC) are metrics used to evaluate the classification results. A 5-fold cross validation was implemented, and the mean area under the curve (AUC) is chosen as the evaluating standard. The result for the dataset generated by Eq. 8 is shown in Fig. 6, whose mean AUC of the ROC curve is $0.82\pm0.07$ and the mean AUC of the PRC is $0.84\pm0.11$. When AUC is significantly above 0.5, the result is also statistically significant, although the actual p-value may be hard to calculate. Hence, we believe our result is sufficient to demonstrate proof-of-principle.

Because normal electric meters represent the majority of day-to-day usage, the proportion of normal meters has been tested from 0.5 to 0.9 to reveal the influence of different proportions of positive samples (normal electric meters) in data generation. Results show that the mean AUC of ROC curve remains stable above 0.7 AUC of the ROC curve despite the proportional changes. Considering real-world applications, the proportion of a positive sample can be higher, which could be a future work after obtaining a larger dataset.

*D. Comparison among Various Classical Methods*

LSTM is compared with several classical methods. The same input data is used and compared with Bayesian ridge regression [31], elastic net [32], gradient boosting regression [33] and LSTM methods on residual error prediction. Figure 5(c) shows the comparison, indicating that the values predicted by the LSTM method have a higher probability of finding inaccurate meters as shown by the pink shading than other methods.

Following the standard in the detection task to check the test value in between LB and UB, a similar standard is made to measure the performance of different regression methods: to count the number of days with predicted $E$ falling below the lower bound (observed value minus threshold) or jumping above the upper bound (observed value plus the threshold). The percentage of counted days to total days is calculated as the measurement, designated as target rate. The result is shown in Table 1, which coincides with the pink shading in Fig. 6. These results show that LSTM and GBR outperformed others in most situations. LSTM performed better than GBR, with a target rate of 95.8% when the threshold equaled 8. A larger threshold is less meaningful, as the accuracy has been high enough (close to 100%). This threshold is henceforth used in the detection procedure.

*E. Analysis of Different Model Configurations*

To compare the combined model mentioned before with the single input models using TS-RP CNN, the AUC of ROC curve for matrix-input and sequence-input was computed. Table 2 shows that AUC is 0.52 when the sequence data is taken as input and AUC can reach 0.82 when both the sequence data and the matrix data are taken as input with pre-training using the ImageNet dataset.

In addition to the VGG-1D CNN combination, other combinations are also tested as components of the dual input network architecture: VGG+BiLSTM [34], ResNet50 [35]+1D-CNN and ResNet50+BiLSTM, as shown in Table 2. These network combinations failed to outperform the VGG+1D-CNN combination in the AUC metric.

## 4. CONCLUSION

In this paper, an approach is presented to detect inaccurate smart meters in residential areas. The TS-RP CNN is proposed for an intelligent system of inaccurate smart meter detection. To the best of our knowledge, this study is the first method to apply deep learning in malfunctional meter detection. The main strategy of this approach is first to judge whether there are any inaccurate meters under the master meter in a residential area using LSTM and then to identify which submeter fails using TS-RP CNN model. It shows that TS-RP CNN successfully identified inaccurate meters inside a target residential area with good accuracy. This paper has also demonstrated that the model reaches equivalent or better performance than several conventional machine learning models. Compared to a single-input model, TS-RP CNN is more comprehensive, where the sequence features in 1D-CNN and the recurrent features (phase-space transformation of sequential data) can assist each other, thereby benefitting classification as a whole.

Detection of inaccurate smart meters is important and valuable for targeted smart meter replacement, especially for some countries that have a great demand for smart meters, but few researchers have worked on this topic. This study provides some insight into the need for more accurate, longer-lasting smart meters, with the hope that more researchers will work on this problem and other problems in the smart meter field. Furthermore, the method is not only useful for increasing the service life span of smart meters, preventing unnecessary replacement, but also provides a general framework for detecting malfunctions in other instruments that generate sequential data, such as charging pile and engine.

There are some limitations of this study. Only the data set from a single location was used in this study. The method needs to be tested using more data. In addition, different types of simulated submeter errors in addition to the linear type used in this study should be explored. Preferably real malfunctional submeters should be used for testing. These represent possible future studies.

REFERENCES

[1] F. Guerrini, "Smart Meters: Between Economic Benefits and Privacy Concerns," *Forbes, June,* vol. 1, 2014.
[2] N. Kshetri, "The evolution of the internet of things industry and market in China: An interplay of institutions, demands and supply," *Telecommunications Policy,* vol. 41, no. 1, pp. 49-67, 2017.
[3] A. T. D. Technologies, "Market Data: Smart Meter Global Forecasts," 2019. [Online]. Available: https://www.navigantresearch.com/reports/market-data-smart-meter-global-forecasts.
[4] R. Silipo and P. Winters, "Big Data Smart Energy and Predictive Analytics," *Time Series Prediction of Smart Energy Data,* vol. 1, p. 37, 2013.
[5] V. Di Cosmo, S. Lyons, and A. Nolan, "Estimating the impact of time-of-use pricing on Irish electricity demand," *The Energy Journal,* pp. 117-136, 2014.
[6] Z. Wu, T. Zhao, L. He, and X. Shen, "Smart grid meter analytics for revenue protection," in *2014 International Conference on Power System Technology*, 2014: IEEE, pp. 782-787.
[7] T. Zufferey, U. Andreas, K. Stephan, and H. Gabriela, "Forecasting of smart meter time series based on neural networks," *International Workshop on Data Analytics for Renewable Energy Integration,* 2016.
[8] W. Sheng, K. Liu, H. Niu, Y. Wang, and J. Zhao, "The anomalous data identification study of reactive power optimization system based on big data," in *Probabilistic Methods Applied to Power Systems (PMAPS), 2016 International Conference on*, 2016: IEEE, pp. 1-5.
[9] N. He, "Research on Big Data Analysis of Smart Meter Fault," *Management & technology of SME,* pp. 142-145, 2016.
[10] J. Yang, P. Jiang, G. Chen, T. Yuan, and F. Xue, "Study on Smart Meters Fault Identification Model," *Electrical Energy Management Technology,* pp. 30-36, 2017.
[11] H. Yao, X. Wang, L. Wu, D. Jiang, T. Luo, and D. Liang, "Prediction Method for Smart Meter Life Based On Big Data," *Procedia Engineering,* vol. 211, pp. 1111-1114, 2018.
[12] K. Gajowniczek and T. Ząbkowski, "Short term electricity forecasting using individual smart meter data," *Procedia Computer Science,* vol. 35, pp. 589-597, 2014.
[13] D. N. Nikovski *et al.*, "Smart meter data analysis for power theft detection," in *International Workshop on Machine Learning and Data Mining in Pattern Recognition*, 2013: Springer, pp. 379-389.
[14] Z. Zhang, Q. He, J. Gao, and M. Ni, "A deep learning approach for detecting traffic accidents from social media data," *Transportation Research part C: emerging technologies,* vol. 86, pp. 580-596, 2018.
[15] T. Kim and S. Cho, "Web traffic anomaly detection using C-LSTM neural networks," *Expert Systems with Applications,* vol. 106, pp. 66-76, 2018.
[16] W. Kong, Z. Y. Dong, Y. Jia, D. J. Hill, Y. Xu, and Y. Zhang, "Short-term residential load forecasting based on LSTM recurrent neural network," *IEEE Transactions on Smart Grid,* vol. 10, no. 1, pp. 841-851, 2017.
[17] D. Monner and J. A. Reggia, "A generalized LSTM-like training algorithm for second-order recurrent neural networks," *Neural Networks,* vol. 25, pp. 70-83, 2012.
[18] S. Hochreiter and J. Schmidhuber, "Long short-term memory," *Neural Computation,* vol. 9, no. 8, pp. 1735-1780, 1997.
[19] K. Greff, R. K. Srivastava, J. Koutník, B. R. Steunebrink, and J. Schmidhuber, "LSTM: A search space odyssey," *IEEE Transactions on Neural Networks Learning Systems,* vol. 28, no. 10, pp. 2222-2232, 2017.
[20] A. Sharif Razavian, H. Azizpour, J. Sullivan, and S. Carlsson, "CNN features off-the-shelf: an astounding baseline for recognition," in *Proceedings of the IEEE conference on computer vision and pattern recognition workshops*, 2014, pp. 806-813.
[21] D. Wang and D. Liu, "MusiteDeep: A deep-learning framework for protein post-translational modification site prediction," in *2017 IEEE International Conference on Bioinformatics and Biomedicine (BIBM)*, 2017: IEEE, pp. 2327-2327.
[22] N. Hatami, Y. Gavet, and J. Debayle, "Classification of time-series images using deep convolutional neural networks," in *Tenth International Conference on Machine Vision (ICMV 2017)*, 2018, vol. 10696: International Society for Optics and Photonics, p. 106960Y.
[23] N. Marwan, "Recurrence Plots and Cross Recurrence Plots," 2016. [Online]. Available: http://www.recurrence-plot.tk/.
[24] N. Marwan, M. C. Romano, M. Thiel, and J. Kurths, "Recurrence plots for the analysis of complex systems," *Physics Reports,* vol. 438, no. 5-6, pp. 237-329, 2007.
[25] E. Bressert, *Scipy and Numpy: an overview for developers*. " O'Reilly Media, Inc.", 2012.
[26] S. J. Pan and Q. Yang, "A survey on transfer learning," *IEEE Transactions on knowledge data engineering,* vol. 22, no. 10, pp. 1345-1359, 2010.
[27] J. Deng, W. Dong, R. Socher, L. Li, K. Li, and F. Li, "Imagenet: A large-scale hierarchical image database," in *2009 IEEE conference on computer vision and pattern recognition*, 2009: IEEE, pp. 248-255.
[28] Y. Han, S. Zhang, and Z. Geng, "Multi-Frequency Decomposition with Fully Convolutional Neural Network for Time Series Classification," in *2018 24th International Conference on Pattern Recognition (ICPR)*, 2018: IEEE, pp. 284-289.
[29] F. Chollet, "Keras Documentation," 2016. [Online]. Available: https://keras.io/.
[30] M. Abadi *et al.*, "Tensorflow: A system for large-scale machine learning," in *12th Symposium on Operating Systems Design and Implementation (16)*, 2016, pp. 265-283.
[31] Q. Shi, M. Abdel-Aty, and J. Lee, "A Bayesian ridge regression analysis of congestion's impact on urban expressway safety," *Accident Analysis Prevention,* vol. 88, pp. 124-137, 2016.
[32] H. Zou and T. Hastie, "Regularization and variable selection via the elastic net," *Journal of the Royal Statistical Society: Series B,* vol. 67, no. 2, pp. 301-320, 2005.
[33] C. Li, "A Gentle Introduction to Gradient Boosting," 2016. [Online]. Available: http://www.chengli.io/tutorials/gradient_boosting.pdf.
[34] Z. Huang, W. Xu, and K. Yu, "Bidirectional LSTM-CRF models for sequence tagging," *arXiv preprint arXiv:1508.01991,* 2015.
[35] K. He, X. Zhang, S. Ren, and J. Sun, "Deep residual learning for image recognition," in *Proceedings of the IEEE conference on computer vision and pattern recognition*, 2016, pp. 770-778.

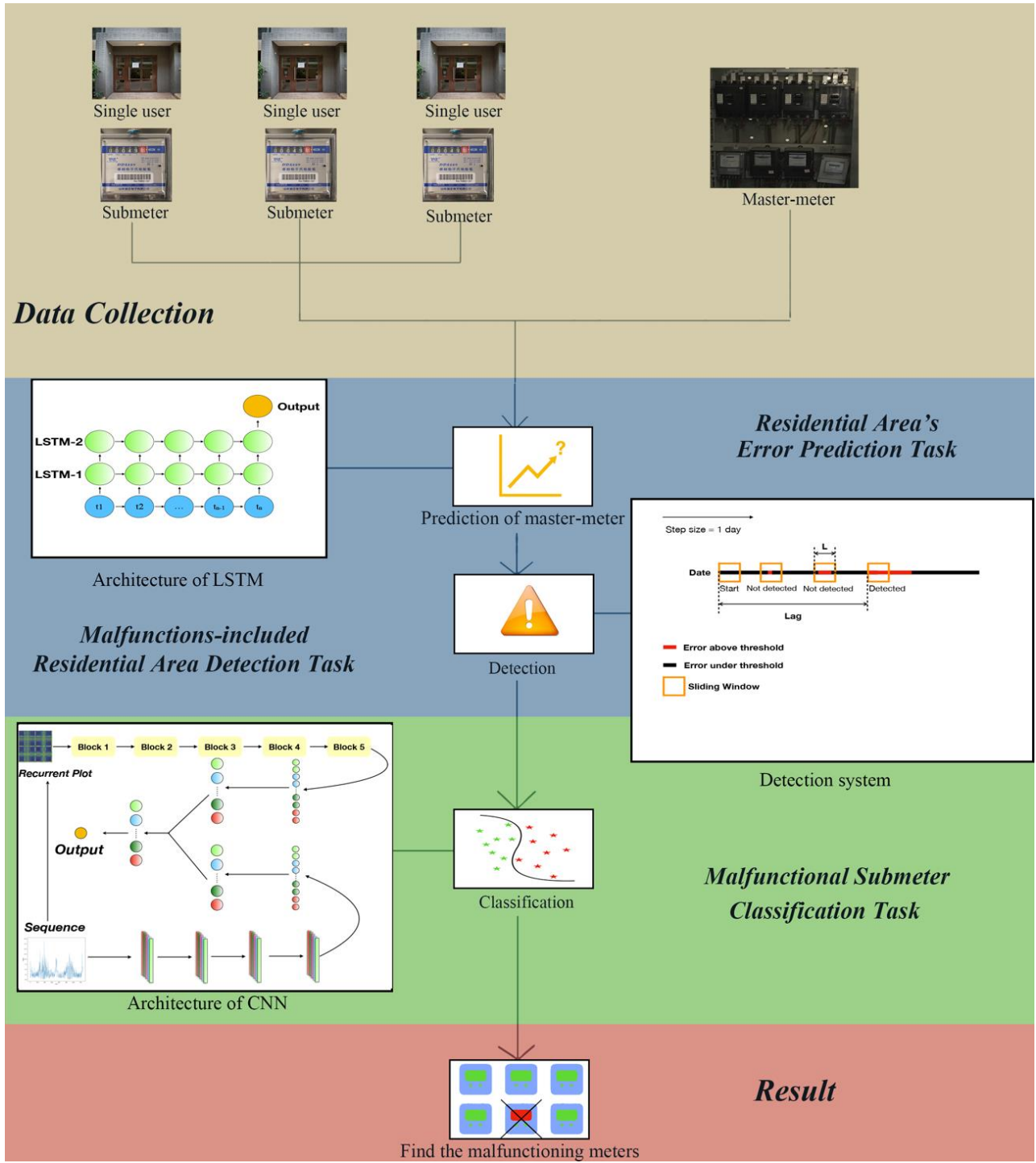

Fig. 1. Workflow of inaccurate smart meter detection. After the data collection, there are three tasks: (1) predicting residual error vs. time based on historical data; (2) embedding a measurement error in some submeters and detecting accurate submeter appears in a residential area based on (1); and (3) following (2) classifying whether each submeter is inaccurate.

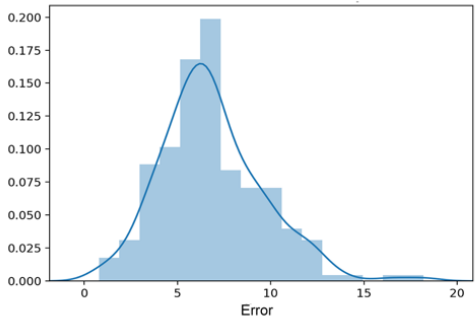

Distribution of errors from January to July

(a)

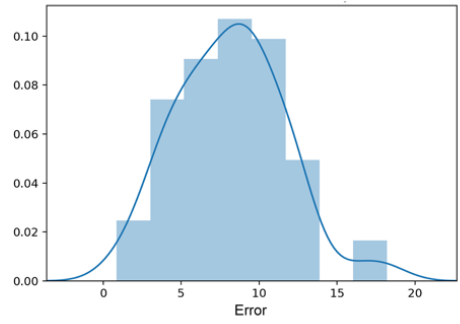

Distribution of errors of January and February

(b)

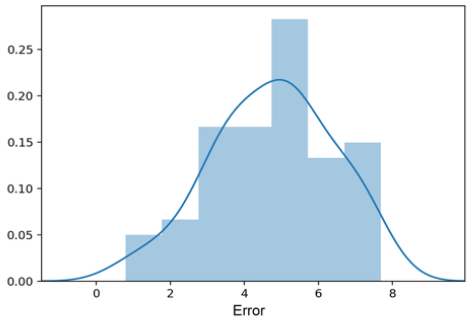

Distribution of errors of March and April

(c)

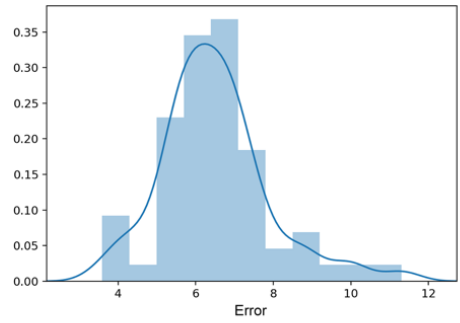

Distribution of errors of May and June

(d)

Fig. 2 Distribution of residual errors between master meters and SSub in different months.

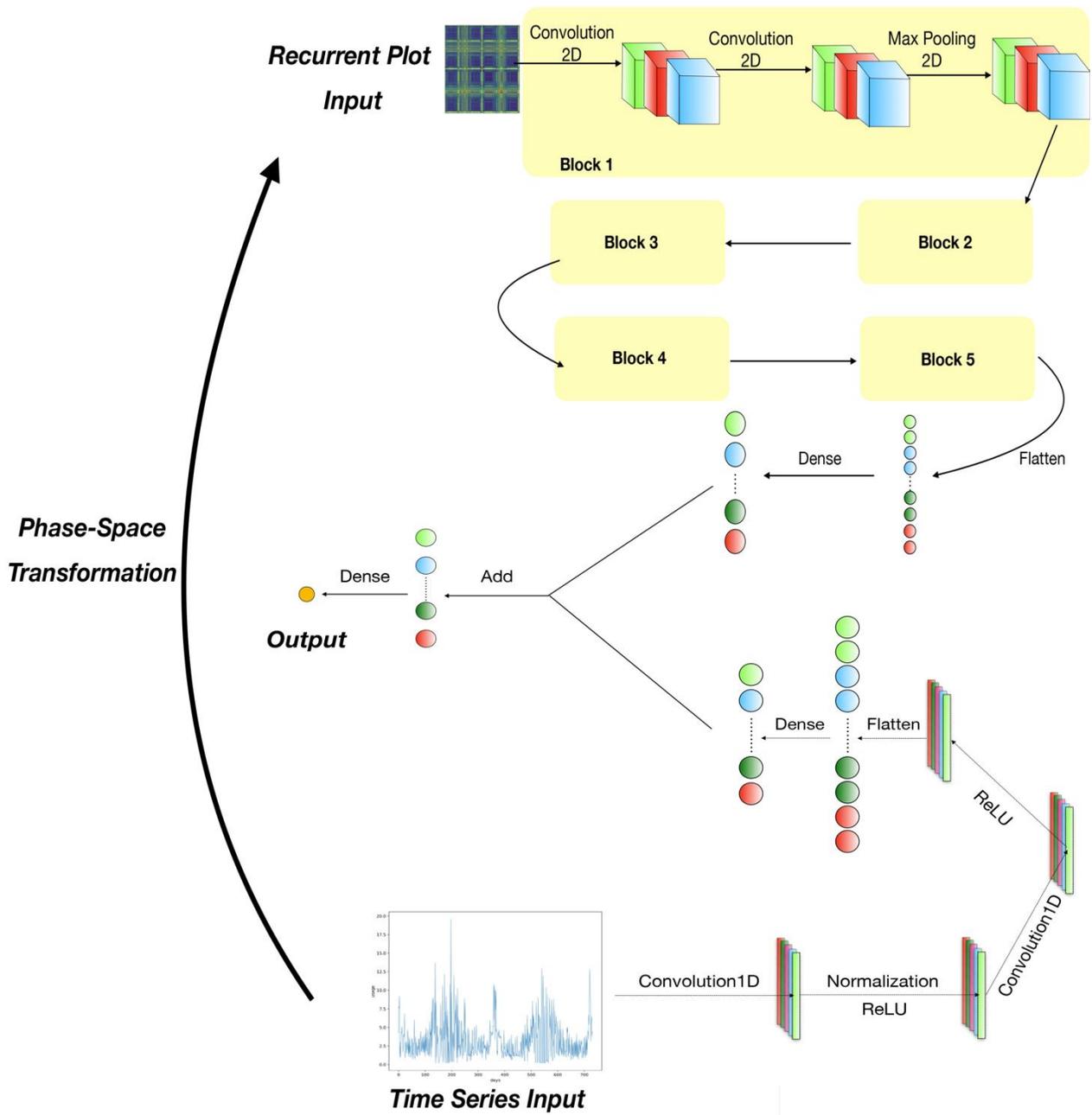

Fig. 3 Time series-recurrence plot (TS-RP) CNN architecture. Firstly, we apply the 1D-CNN architecture to classify the time series input. After the phase-space transformation, the recurrent plot input is created and applied in 2D-CNN. Then, the two types of input branches are merged and applied in the TS-RP CNN architecture to get the output. The colors in the figure are used to illustrate various components.

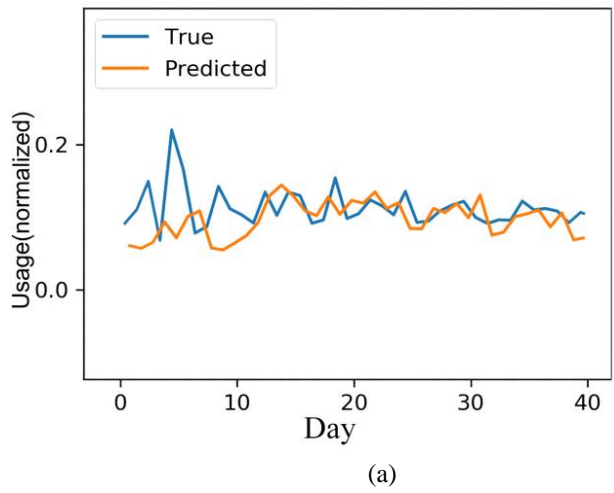 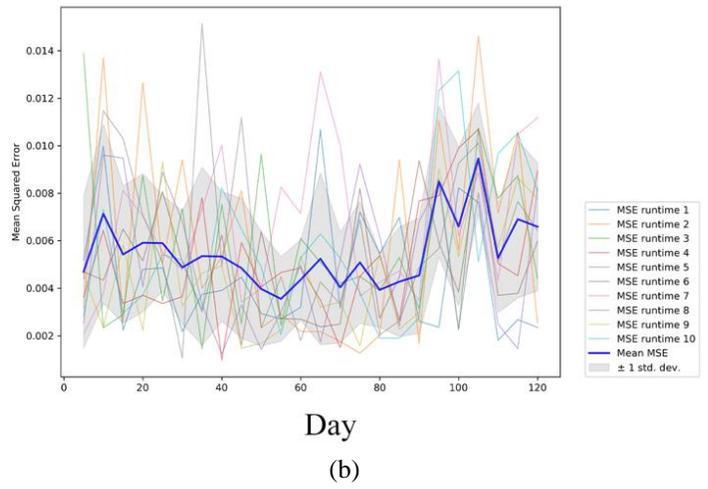

(a)                                         (b)

Fig. 4 (a) Observed vs. predicted user usage with respect to time, and (b), exploring the impact of different window sizes in days based on mean square error (MSE) of the prediction, averaged over the 10.

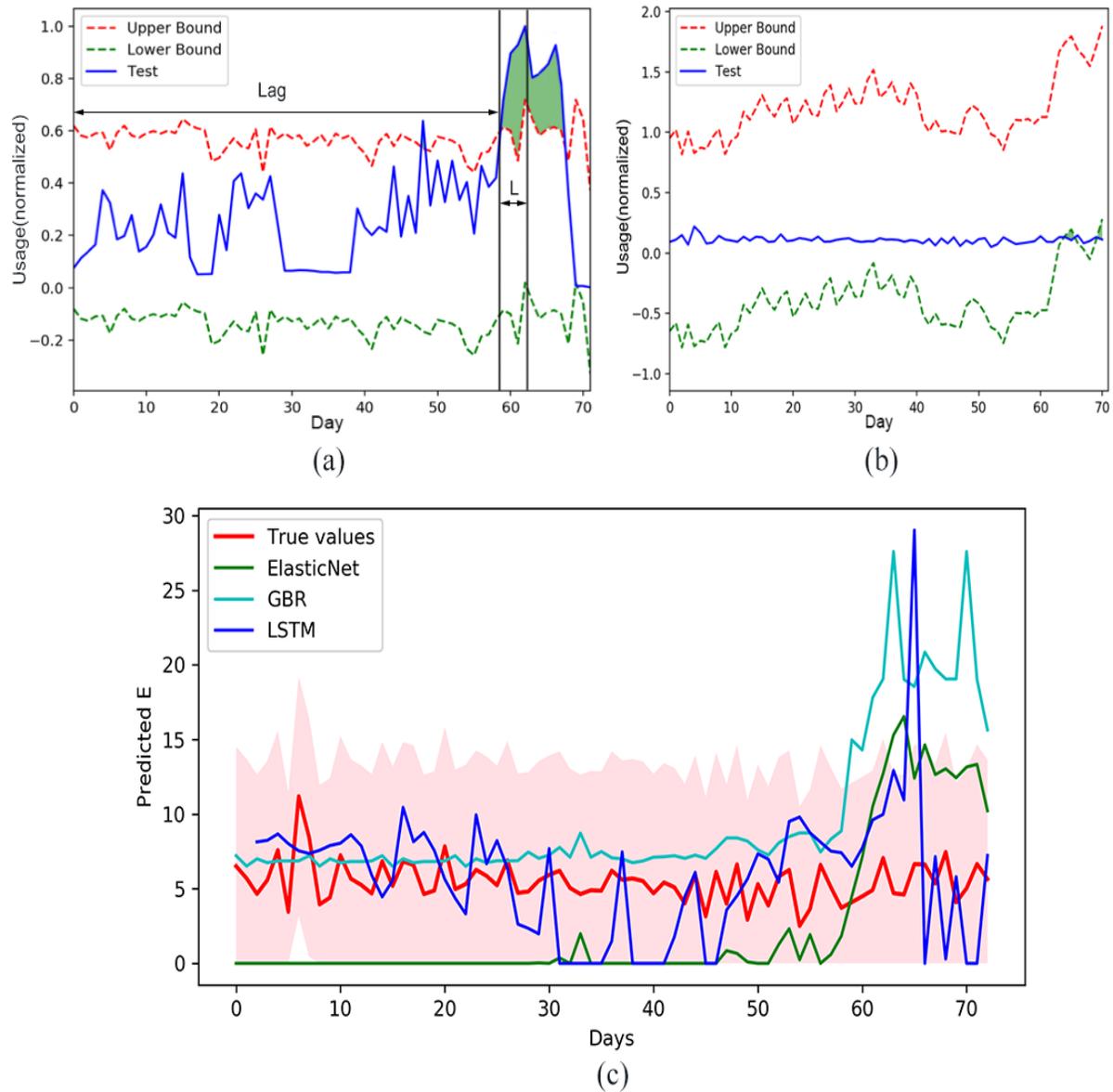

Fig. 5 (a) A detection example of a residential area with malfunction. When the threshold is 0.5 and the window size is 4, the inaccurate submeters in that residential area are detected. (b) shows that another residential area with the same parameter setting is not detected as including inaccurate meters. (c) shows a comparison of regression result on the residual error prediction task when the threshold is 8. $E$ is defined in Equation 1. The red line represents the observed value of the measurement residual error, and other lines are the predicted values of the measurement residual error of using Bayesian Ridge, Elastic Net, Gradient Boosting Regression, and LSTM. The pink area indicates that the difference to the observed value is less than the threshold. The predicted values falling within this area are considered without inaccurate submeters.

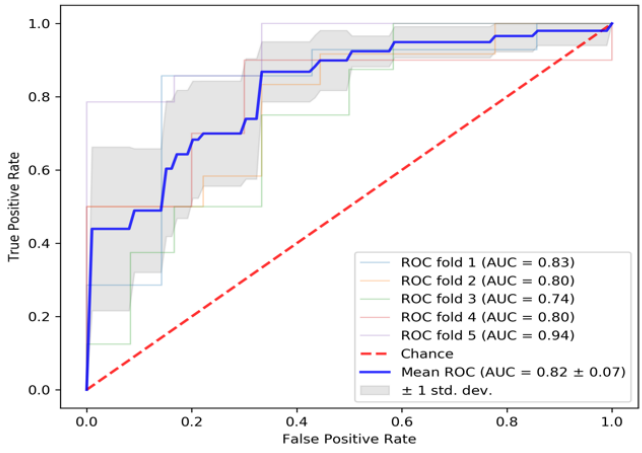 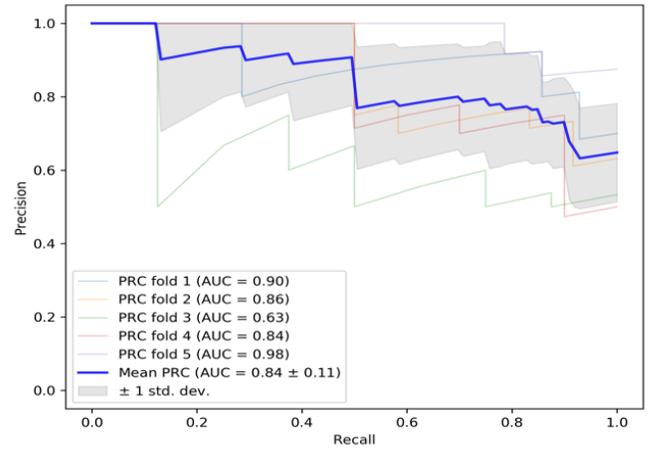

Fig. 6 ROC curve and PRC of the classification of inaccurate smart meters in a 5-fold cross-validation.

Table 1. Evaluation of the prediction of classical methods and LSTM

| Threshold | Classical Methods | | LSTM |
|---|---|---|---|
| | Elastic Net | GBR | |
| 0.5 | 1 (1.4%) | 5 (6.9%) | **5 (6.9%)** |
| 1 | 1 (1.4%) | 13 (18.1%) | **17 (23.6%)** |
| 4 | 13 (18.1%) | **52 (72.2%)** | 42 (58.3%) |
| 6 | 52 (72.2%) | 58 (80.6%) | **65 (90.3%)** |
| 8 | 66 (91.7%) | 59 (81.9%) | **69 (95.8%)** |

The numbers show the number of days with predicted $E$ falling below the lower bound or jumping above the upper bound, and the percentage of counted days to total days as the measurement (target rate, in parentheses)

Table 2. Comparison of different network architecture combinations

| AUC of ROC curve | Single-input Modeling | | Dual-input Modeling | | | |
|---|---|---|---|---|---|---|
| Fold number | Sequence-input | Matrix-input* | TS-RP CNN (VGG*+1D-CNN) | VGG* + BiLSTM* | ResNet50* + 1D-CNN | ResNet50*+BiLSTM |
| Fold 1 | 0.29 | 0.47 | 0.83 | 0.91 | 0.65 | 0.37 |
| Fold 2 | 0.62 | 0.71 | 0.80 | 0.75 | 0.58 | 0.45 |
| Fold 3 | 0.46 | 0.17 | 0.74 | 0.71 | 0.43 | 0.54 |
| Fold 4 | 0.68 | 0.41 | 0.80 | 0.80 | 0.57 | 0.52 |
| Fold 5 | 0.55 | 0.79 | 0.94 | 0.87 | 0.30 | 0.48 |
| Mean (±1 std.) | 0.52±0.14 | 0.51±0.22 | 0.82±0.07 | 0.81±0.07 | 0.51±0.13 | 0.60±0.13 |

\* performed with pre-training using the ImageNet dataset.